\documentclass[preprint,12pt,showpacs,showkeys]{revtex4-1}

\usepackage[T1]{fontenc} % if needed

\usepackage[brazil, english]{babel}
\usepackage[utf8]{inputenc}
\usepackage{graphicx}
\usepackage{epsfig}
\usepackage{slashed}
\usepackage{amssymb}
\usepackage{amsmath}
\usepackage[titletoc]{appendix}
\usepackage{color}
\usepackage{soul}

\begin{document}

\title{Hadronic Spectra from Deformed AdS Backgrounds}

\author{Eduardo Folco Capossoli$^{1,2}$}
\email{eduardo_capossoli@cp2.g12.br} 
\author{Miguel Angel Martín Contreras$^3$}
\email{miguelangel.martin@uv.cl}
\author{Danning Li$^4$}
\email{lidanning@jnu.edu.cn}
\author{Alfredo Vega$^3$}
\email{alfredo.vega@uv.cl} 
\author{Henrique Boschi-Filho$^{1}$}
\email{boschi@if.ufrj.br}  
\affiliation{$^1$Instituto de F\'\i sica, Universidade Federal do Rio de Janeiro, 21.941-972 - Rio de Janeiro-RJ - Brazil \\
 $^2$Departamento de F\'\i sica / Mestrado Profissional em Práticas da Educação Básica (MPPEB), Col\'egio Pedro II, 20.921-903 - Rio de Janeiro-RJ - Brazil\\
 $^3$Instituto de Física y Astronomía, Universidad de Valpara\'iso, A. Gran Breta\~na 1111, Valpara\'iso, Chile\\ 
$^4$Department of Physics and Siyuan Laboratory, Jinan University, Guangzhou 510632, China}

\begin{abstract} 

Because of the presence of modified warp factors in metric tensors, we use deformed AdS$_5$ spaces to apply the AdS/CFT correspondence  to calculate the spectra for even and odd glueballs, scalar and vector mesons, and baryons with different spins. For the glueball cases, we derive their Regge trajectories and compare them with those related to the pomeron and the odderon. For the scalar and vector mesons as well as baryons the determined masses are compatible with the PDG. In particular for these hadrons we found Regge trajectories compatible with another holographic approach as well as with the hadronic spectroscopy, which present an universal Regge slope of approximately 1.1 GeV$^2$.
\end{abstract}

\keywords{hadronic spectra, AdS/QCD model, Regge trajectories}
%\pacs{11.25.Wx, 11.25.Tq, 12.38.Aw, 12.40.Yx}

\maketitle
%\flushbottom

\section{Introduction}

Quantum Chromodynamics (QCD) is a non-Abelian quantum field theory employed for dealing with strong interactions. Although its boasts enormous success in the high energy regime, the use of QCD is difficult when investigating processes that occur at low energies (IR regions) because of the failure of the perturbative approach. This peculiar feature of the QCD is related to the fact that it is a confining theory in the IR, implying that only bound states of quarks or gluons are observed. 

Hadronic spectroscopy is a highly interesting field with regard to the application of new approaches to extract information about hadronic properties, given that results are comparable with the experimental data. 

Among several techniques to handle within the field of Hadronic spectroscopy, there is one that emerged in 1997 proposed by Juan Maldacena, referred to as the  Anti de Sitter/Conformal Field Theory or AdS/CFT correspondence \cite{Maldacena:1997re, Gubser:1998bc,Witten:1998qj,Witten:1998zw,Aharony:1999ti}. This correspondence is very useful, as it provides guidance on how to relate a weak coupling theory, which is in this case represented by a superstring theory in a ten-dimensional curved space, named $AdS_5 \times S^5$ with a strong coupling theory which is a super conformal Yang-Mills theory with extended supersymmetry ${\cal N} = 4$, symmetry group $SU(N \to \infty)$ in a flat four-dimensional Minkowski space.

However, the AdS/CFT correspondence cannot be used directly to reproduce QCD, as the latter is not a conformal theory, as it possesses numerous different scales (masses, critical temperature, etc.). 
v
Some proposals appeared to break the conformal invariance and build effective theories known as  AdS/QCD models, e.g, the hardwall model. In this model, the conformal symmetry is broken via introduction of a hard IR cutoff at a certain value $z_{max}$ of the holographic coordinate $z$ and by considering only a slice of the $AdS_5$ space within the interval $[0, z_{max}]$ \cite{Polchinski:2001tt, BoschiFilho:2002vd, BoschiFilho:2002ta}. Achievements in hadronic spectroscopy within the hardwall are presented in several studies \cite{Erlich:2005qh, DaRold:2005mxj, deTeramond:2005su, DaRold:2005vr, Pomarol:2008aa, Wang:2009wx, Li:2013lfa}.

Another example of breaking the conformal invariance is given by the softwall model. In this model a soft IR cutoff via an introduction of a dilaton field in the action. This approach was proposed in \cite{Karch:2006pv} to study mesonic spectroscopy. Usually, this model is referred to as the original softwall model. Several modifications of this model were considered subsequently to deal with hadronic spectroscopy as presented in e.g., Refs. \cite{Huang:2007fv, Vega:2008te, Branz:2010ub, Gutsche:2011vb, Afonin:2012jn, Fang:2016uer, Cortes:2017lgz, Contreras:2018hbi, Afonin:2018era, Gutsche:2019blp}. Further addressing some modification in the Refs.\cite{Andreev:2006vy, Andreev:2006ct, Forkel:2007cm,  White:2007tu, Bruni:2018dqm, Rinaldi:2017wdn}, instead of the introduction of a dilation in the action, a modified warp factor in the AdS metric was considered. Particularly, in Ref. \cite{Forkel:2007cm} such a modification was proposed to study hadronic spectroscopy. Other modifications of the softwall model were used in Refs. \cite{Andreev:2006ct, White:2007tu, Bruni:2018dqm} to discuss the quark-antiquark potential and in Ref. \cite{Rinaldi:2017wdn} to deal with scalar and tensor glueballs. One open problem associated with the softwall model is the sign of the dilaton. In the original case, the dilaton is an exponential with a negative argument \cite{Karch:2006pv}. In Refs. \cite{deTeramond:2009xk, Zuo:2009dz, Nicotri:2010at}, authors argued that a positive dilaton is preferred, which the authors of Ref. \cite{Karch:2010eg} disagree with. These authors also point out that a positive dilaton implies the existence of a massless scalar in the spectrum. 

In this study, inspired by Refs. \cite{Andreev:2006vy, Andreev:2006ct, Forkel:2007cm}, we investigate these problems with modified warp factors in the $AdS_5$ metric instead of introducing dilaton fields in the action. In this sense, in our set-up we consider deformed AdS backgrounds. 
Subsequently, using this approach, we compute the hadronic spectra for several particles with different spins. We employ the same form for the warp factor in the metric by fitting the free parameter in each case. The values of the parameters are observed to be different for each sector. This scenario is similar to the case of the original softwall model, where different dilaton fields are needed for each particle sector. The main advantage of our approach is that we can also directly deal with fermions, contrary to the original softwall model. Furthermore, our approach provides appropriate masses and Regge trajectories, for instance, for odd and even spin glueballs. 

This paper is organized as follows. In Section \ref{sec2} we present a brief review of the original softwall model and our deformed AdS background. In Section \ref{gb}, we apply our model to the even and odd spin glueball states.  
In Section \ref{meson}, we study the case of scalar mesons obtaining their spectra. 
We  calculate the hadronic spectra for the vector mesons in Section \ref{sec4} and in Section \ref{hf} we address the baryonic case with spins 1/2, 3/2 and 5/2. For those particles, we also obtain the corresponding Regge trajectories. In particular, for the glueballs we derive the Regge trajectories related to the pomeron and the odderon. 
Finally, in Section \ref{sec6} we  present the conclusions and final comments.

%%%%%%%%%%%%%%%%%%%%%%%
%%%%%%%%%%%%%%%%%%%%%%%
%%%%%%%%%%%%%%%%%%%%%%%

\section{Softwall Model and Deformed AdS Set-Up}\label{sec2}
There are at least two interesting reasons behind the emergence of the softwall model. The first is related to the introduction of the soft IR cutoff instead a hard cutoff as in the hardwall model, as this approach seems more natural. The second reason lies in the fact that the softwall model truly yields linear Regge trajectories, which was an established behavior since the beginning of hadronic spectroscopy, so that
\begin{equation}
J(m) \approx \alpha' m^2 + \alpha_0,
\end{equation}
\noindent where $J$ is the total angular momentum; $m$ represents the hadronic mass; $\alpha'$ (Regge slope) and $\alpha_0$ are constants. The relationship between  radial excitation $n$ and its squared hadron mass, given by:
\begin{equation}
m^2 \approx \beta' n + \beta_0, 
\end{equation}
\noindent with $\beta'$ and $\beta_0$ as constants.

In the original formulation of the softwall model, the action of the fields, up to some constant, is described by:
\begin{equation}\label{acao_sw}
S = \int d^5 x \sqrt{-g} \; e^{-\Phi(z)} {\cal L}, 
\end{equation}
\noindent where $\Phi(z)$ is the dilaton field, usually given by $\Phi(z)=  kz^2$, where $|k| \sim \Lambda^2_{QCD}$, and ${\cal L}$ is the Lagrangian density.

The main difference between the original softwall model and the present study is the modified $AdS_5$ metric tensor using an exponential warp factor for all glueballs and hadrons. In Ref. \cite{Forkel:2007cm} the authors used different warp factor profiles, usually logarithmic ones, for each hadronic sector. 

As we employ the same warp factor profile in the AdS space for all glueballs and hadrons, we refer the approach of this study as a deformed $AdS_5$ background.
Then, we write the deformed $AdS_5$ metric as: 
\begin{equation}\label{gs}
ds^2 = g_{mn} dx^{m} dx^{n}= \frac{R^2}{z^2}e^{kz^2}(dz^2 + \eta_{\mu \nu}dx^{\mu} dx^{\nu}) = e^{2 A(z)}(dz^2 + \eta_{\mu \nu}dx^{\mu} dx^{\nu}), 
\end{equation}
\noindent where $R$ is the usual AdS radius (from here onward, we assume $R=1$ throughout this text), $\eta_{\mu \nu}$ is the flat Minkowski space metric tensor in four dimensions with signature $(-, +, +, +)$, $z$ is the holographic coordinate, and $x^{m} = (z, x^{\mu})$ for $\mu = 0,\cdots , 3$. The warp factor $A(z)$ in Eq. \eqref{gs} can be read as:
\begin{equation}\label{az}
 A(z) =  -\log(z) + \frac{kz^2 }{2}\,. 
 \end{equation} 

In our model, the action for the fields is given as:
 \begin{equation}\label{acao_swd}
S = \int d^5 x \sqrt{-g} \; {\cal L}, 
\end{equation}
\noindent where $g$ is the determinant of the five-dimensional metric tensor presented in Eq. \eqref{gs}.
 
\section{Hadronic Spectra for glueballs states}\label{gb}

Fritzsch and Gell-Mann pointed out in Refs \cite{Dobbs:2015dwa, Fritzsch:1972jv}.  ``If the quark-gluon field theory indeed yields a correct description of strong interactions, there must exist glue states in the hadron spectrum”. This sentence does really reveals the importance of those ``glue states'' nowadays referred to as glueballs. Glueballs are colorless bound states of gluons predicted by QCD but not experimentally detected to date. 

Glueballs are characterized by $J^{PC}$ where $J$ (even or odd) is the total angular momentum, $P$ is the $P-$parity (spatial inversion) and $C = $  is  the $C-$parity (charge conjugation) eigenvalues. For the glueballs case, $P=(-1)^L $ and $C=(-1)^{L+S}$. 

Numerous experimental efforts were conducted in the search for glueballs \cite{Haguenauer:1993kan, Avila:2006wy, Ablikim:2006db, Bai:2003ww}. Some theoretical and non-holographic approaches are described in Refs. \cite{ Morningstar:1999rf, Meyer:2004jc, Chen:2005mg, Lucini:2001ej, Szczepaniak:2003mr, Mathieu:2008bf}. The holographic approach is presented in Refs. \cite{BoschiFilho:2005yh, Colangelo:2007pt, Capossoli:2013kb, Rodrigues:2016cdb, BoschiFilho:2012xr, Li:2013oda, Capossoli:2015ywa, Capossoli:2016kcr, Capossoli:2016ydo, FolcoCapossoli:2016ejd}.

In this study based on a deformed AdS space, we compute the masses of even spin glueballs with $P=C=+1$ and odd spin glueballs with $P=C=-1$. Even spin glueballs with $P=C=+1$ are particularly interesting, as in the Chew-Frautschi plane, their states lie on the Pomeron Regge trajectory. In contrast, odd spin glueballs with $P=C=-1$ lie on the odderon Regge trajectory.

We start our calculation using the standard action for a massive scalar field $X$ in $5D$ space, given by:
\begin{equation}\label{esc_sw}
S = \int d^5 x \sqrt{-g}\; [ g^{mn} \partial_m X \partial^n X + M_5^2 X^2 ].
\end{equation} 
From the action \eqref{esc_sw} one can find the following equations of motion, so that:
\begin{equation} \label{eom_esc}
\partial_m[\sqrt{-g} g^{mn} \partial_n X] - \sqrt{-g}M_5^2 X = 0, 
\end{equation}
where $g^{mn} = e^{-2A(z)} \eta^{m n}$.

The Eq. \eqref{eom_esc} can be written as:
\begin{equation} \label{eom_esc_2}
\partial_m[e^{3A(z)} \eta^{mn} \partial_n X] - e^{5A(z)} M_5^2 X = 0, 
\end{equation}
\noindent with the warp factor $A(z)$ given in Eq. \eqref{az}.

Defining $B(z) = - 3A(z)$, we obtain:
\begin{equation}\label{eomsw2}
\partial_m[e^{-B(z)} ~ \eta^{mn}  \partial_n X] - e^{\frac{-5 B(z)}{3}} M^2_5 X = 0.
\end{equation}
Next, we use a plane wave ansatz with the amplitude only depending on the $z$ coordinate and propagating in the transverse coordinates $x^{\mu}$ with momentum $q_{\mu}$, 
\begin{equation}\label{an}
X (z, x^{\mu}) = v(z) e^{i q_{\mu} x^{\mu}}.
\end{equation}

\noindent After some algebraic manipulation and defining 
$v(z) = \psi (z) e^{\frac{B(z)}{2}}$ we obtain a ``Schr\"odinger-like'' equation:
\begin{equation}\label{eq_4}
- \psi''(z) + \left[ \frac{B'^2(z)}{4}  - \frac{B''(z)}{2} + e^{\frac{-2 B(z)}{3}} M^2_5 \right] \psi(z) = - q^2 \psi(z), 
\end{equation}
\noindent with $B(z) = - 3 A(z)$  and $E = -q^2$ as eigenenergies.

\subsection{Results for even and odd spin glueball spectra}\label{sub}

To compute the glueball masses, Eq. \eqref{eq_4} must be solved numerically. To this end, from the AdS/CFT dictionary we first relate the masses of supergravity fields in the AdS space $(M_5)$ with the scaling dimensions of an operator in the boundary theory $(\Delta)$, such that:
\begin{equation}\label{brod}
M^2_5 = (\Delta - p) (\Delta + p - 4), 
\end{equation}
\noindent where $p$ is the index of a $p-$form. For the case of the scalar glueball $0^{++}$ we obtain $p=0$. Because the scalar glueball is dual to the fields with $M_5 = 0$, its conformal dimension is $\Delta = 4$.

Second, the scalar glueball state is represented on the boundary theory by the operator ${\cal O}_4$, given by:
\begin{equation}\label{fmn}
{\cal O}_4 = \text{Tr}\,\left(F^2\right) = \text{Tr}\,\left( F^{\mu\nu}F_{\mu \nu} \right).
\end{equation}

To raise the total angular momentum $J$, we follow Ref. \cite{deTeramond:2005su} by inserting  symmetrised covariant derivatives in a given operator with spin $S$, such that the total angular momentum after the insertion becomes $S+J$. In the particular case of the operator ${\cal O}_4 = \,\text{Tr} \,F^2$, we obtain:
\begin{equation}\label{4+J}
{\cal O}_{4 + J} = \,\text{Tr}\, \left(FD_{\lbrace\mu1 \cdots} D_{\mu J \rbrace}F\right),
\end{equation}
\noindent with the conformal dimension $\Delta = 4 + J$. For $J=0$ we recover $\Delta = 4$.

Thus, for even spin glueball states after the insertion of symmetrized covariant derivatives, we obtain:
\begin{equation}\label{hsp}
M^2_{5} = J(J+4)\,; \qquad ({\rm even}\, J)\;. 
\end{equation}
Hence, we write Eq.\eqref{eq_4} as:
\begin{equation}\label{eq_4_je}
- \psi''(z) + \left[ \frac{B'^2(z)}{4}  - \frac{B''(z)}{2} + e^{\frac{-2 B(z)}{3}} J(J+4) \right] \psi(z) = - q^2 \psi(z).
\end{equation} 
%\clearpage
Solving Eq.\eqref{eq_4_je}, for even glueball states, one obtains the four-dimensional masses presented in Table \ref{t1}. %%%%%%%%%%%%%%%%%%%%%
\begin{table}
%\begin{ruledtabular}
%\vspace{0.5 cm}
\centering
\begin{tabular}{|c|c|c|c|c|c|c|}
\hline
 &  \multicolumn{6}{c|}{Even glueball states $J^{PC}$}  \\  
\cline{2-7}
 & $0^{++}$ & $2^{++} $ & $4^{++}$ & $6^{++}$ & $8^{++}$ & $10^{++}$  \\
\hline \hline
Masses                                   
&\, 0.76\, &\, 2.08 \,&\, 3.17 \,& \, 4.22 \, &\, 5.26 &\, 6.30\\ \hline
\end{tabular}
\caption{Even glueball masses expressed in {\rm GeV} from Eq. \eqref{eq_4_je} with warp factor constant $k$ as $k_{\rm gbe}= 0.31^2$ {\rm GeV}$^2$.}
%\end{ruledtabular}
\label{t1}
\end{table}
%%%%%%%%%%%%%%%%%%%%%%%%%%%%%%%%%%

\begin{figure}
	\centering
	\includegraphics[scale=0.6]{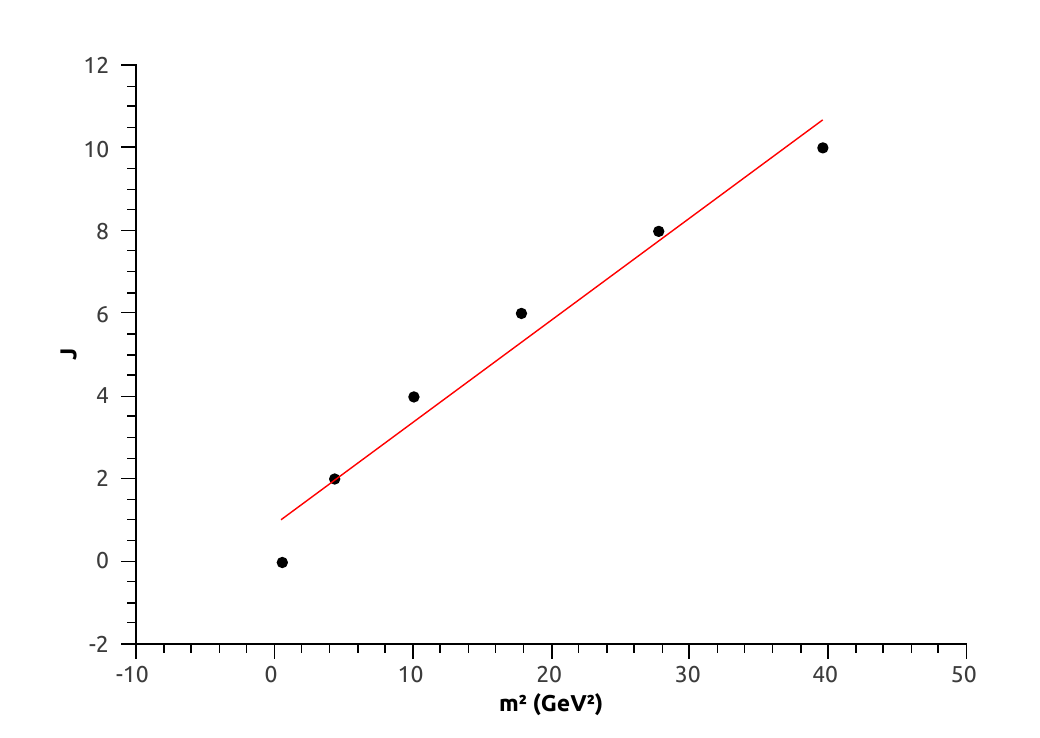}
	\caption{(color online) Approximate linear Regge trajectory associated with the pomeron from Eq. \eqref{rtp}. The dots correspond to the masses found in Table \ref{t1} for even glueball states within the deformed $AdS_5$ space approach .}
	\label{pom}
\end{figure}

Using the data from Table \ref{t1}, we plotted a Chew-Frautschi plane, here represented as $ m^2 \times J $, where $J$ is total angular momentum, and $m^2$ is the squared even glueball mass represented by the dots in figure \ref{pom}. Using a standard linear regression method, we obtain the equation
\begin{equation}\label{rtp}
J(m^2) \approx (0.25 \pm 0.02) m^2 + (0.88 \pm 0.51),
\end{equation}
which represents an approximate linear Regge trajectory associated with the pomeron in agreement with Refs. \cite{Donnachie:1985iz, Donnachie:2002en}.
%%%%%%%%%
%\clearpage

In contrast, for odd glueball states, the operator ${\cal O}_6$ that describes the glueball state $1^{--}$ is given by \cite{Wang:2009wx, Capossoli:2013kb, Csaki:1998qr, Brower:2000rp}:
\begin{equation}
 {\cal O}_{6} =\text{Sym}\,\text{Tr}\left( {\tilde{F}_{\mu \nu}}F^2\right),
 \end{equation} 

\noindent where this dual operator creates odd glueball states at the boundary. This operator has the conformal dimension $\Delta = 6$, and after the insertion of symmetrized covariant derivatives, we obtain:
\begin{equation}\label{6+J}
{\cal O}_{6 + J} = \text{Sym} \,\text{Tr}\left( {\tilde{F}_{\mu \nu}}F D_{\lbrace\mu1 \cdots} D_{\mu J \rbrace}F\right),
\end{equation}

\noindent with $\Delta = 6 + J$. Therefore, 
\begin{equation}\label{hsi}
M^2_{5} = (J+6)(J+2)\,; \qquad ({\rm odd}\, J)\,,
\end{equation}
and we can rewrite Eq.\eqref{eq_4} as:
\begin{equation}\label{eq_4_jo}
- \psi''(z) + \left[ \frac{B'^2(z)}{4}  - \frac{B''(z)}{2} + e^{\frac{-2 B(z)}{3}} (J+6)(J+2) \right] \psi(z) = - q^2 \psi(z).
\end{equation}
Solving Eq.\eqref{eq_4_jo} for odd glueball states, we obtain the four-dimensional masses presented in Table \ref{t2}. 
%%%%%%%%%%%%%%%%%%%%%
\begin{table}
%\begin{ruledtabular}
%\vspace{0.5 cm}
\centering
\begin{tabular}{|c|c|c|c|c|c|c|}
\hline
 &  \multicolumn{6}{c|}{Odd glueball states $J^{PC}$}  \\  
\cline{2-7}
 & $1^{--}$ & $3^{--} $ & $5^{--}$ & $7^{--}$ & $9^{--}$ & $11^{--}$  \\
\hline \hline
Masses                                   
&\, 2.63\, &\, 3.70 \,&\, 4.74 \,& \, 5.78 \, &\, 6.81&\, 7.84\\ \hline
\end{tabular}
\caption{ Odd spin glueball masses expressed in {\rm GeV} as from Eq.\eqref{eq_4_jo} with the warp factor constant $k$ as $k_{\rm gbo}= 0.31^2$ {\rm GeV}$^2$.} 
%\end{ruledtabular}
\label{t2}
\end{table}
%%%%%%%%%%%%%%%%%%%%%%%%%%%%%%%%%%

\begin{figure}
	\centering
	\includegraphics[scale=0.6]{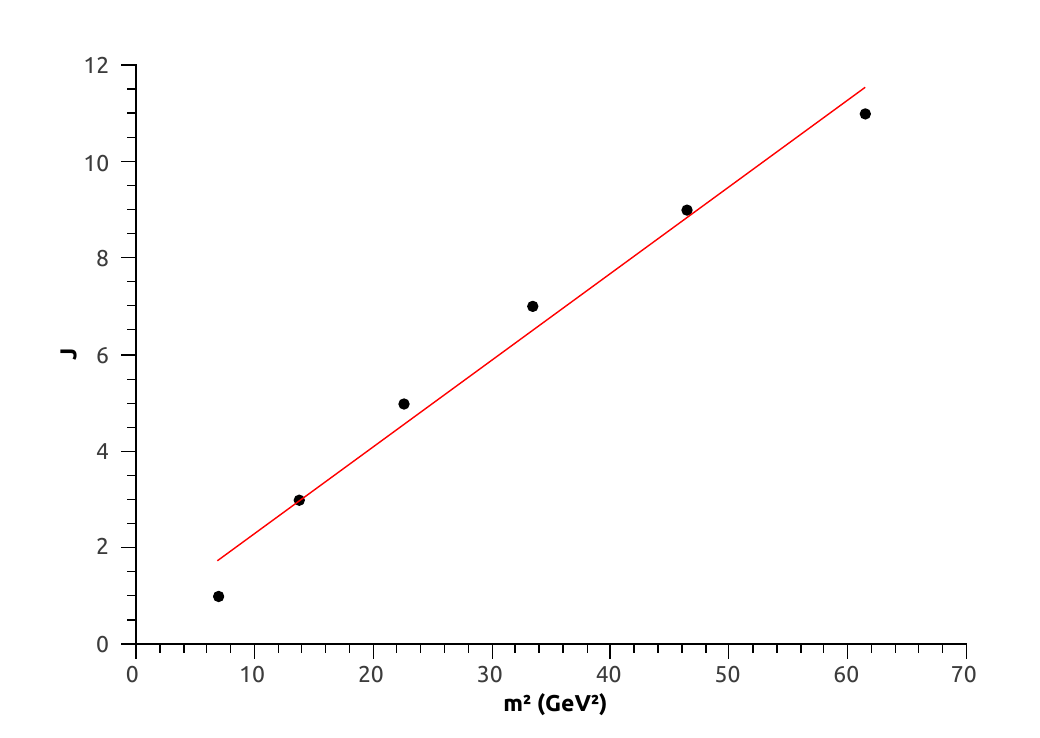}
	\caption{(color online) Approximate linear Regge trajectory associated with the odderon from Eq. \eqref{rto}. The dots correspond to the masses found in Table \ref{t2} within the deformed $AdS_5$ space approach for odd spin glueballs.}
	\label{odd}
\end{figure}

Using the data in Table \ref{t2}, we plotted a Chew-Frautschi plane $ m^2 \times J $ in figure \ref{odd} for odd spin glueballs. Using a standard linear regression method, we obtain the equation
\begin{equation}\label{rto}
J(m^2) \approx (0.18 \pm 0.01) m^2 + (0.47 \pm 0.45)\,, 
\end{equation}
which is in agreement with Ref. \cite{LlanesEstrada:2005jf}, within the nonrelativistic constituent model.

Notably, the value for the constant $k$ in the warp factor $A(z)$ for even spin glueball represented by $k_{\rm gbe}$ and for odd spin glueball represented by $k_{\rm gbo}$ have the same numerical value $k_{\rm gbe} = k_{\rm gbo} = 0.31^2$ {\rm GeV}$^2$. 

To facilitate the comparison between our results with the deformed AdS model and other approaches, we summarize several results provided by the literature in Tables \ref{pomlit} and \ref{oddlit}.

\begin{table}[!h]
\centering
\begin{tabular}{|c|c|c|c|c|}
\hline
 Models used &  \multicolumn{4}{c|}{  Even Glueball States $J^{PC}$} \\  
\cline{2-5}
 & $0^{ + + }$ &  $2^{+ + }$ &  $4^{+ + }$ &  $6^{+ + }$  \\
\hline \hline
 $N_c = 3$ lattice \cite{Meyer:2004gx}                                   
&\,1.475(30)(65)\, &\,  2.150(30)(100) \,&\, 3.640(90)(160) \,& \, 4.360(260)(200) \,  \, \\ \hline
$N_c = 3$ anisotropic lattice  \cite{Morningstar:1999rf}                             
&  1.730(50)(80) &  2.400(25)(120) &  &         \\ \hline
 $N_c = 3$ anisotropic lattice  \cite{Chen:2005mg}           
& 1.710(50)(80) &  2.390(30)(120) &  &        \\ \hline
 $N_c = 3$ lattice \cite{Lucini:2001ej}     
&  1.58(11) &  &  &        \\ \hline
 $N_C \to \infty$ lattice  \cite{Lucini:2001ej}                           
&  1.48(07) &  &      &              \\ \hline
 Constituent models  \cite{Szczepaniak:2003mr}
&  &  2.42 &  2.59 &         \\ \hline
 Constituent models \cite{Mathieu:2008bf}                       
&  &  3.99 &  3.77 &  4.60    \\ \hline 
\end{tabular}
\caption{Glueball masses for $J^{PC}$ states expressed in GeV, with even $J$, achieved with non-holographic models from the literature. Numbers in parentheses represent uncertainties.}
%\end{ruledtabular}
\label{pomlit}
\end{table}
%%%%%%%%%%%%%%%%%%%%%%%%%%%%%%%%%%%%%%%

%%
\begin{table}[h]
\centering
\begin{tabular}{|c|c|c|c|c|}
\hline
 Models used &  \multicolumn{4}{c|}{Odd  glueball states $J^{PC}$}  \\  
\cline{2-5}
 &\qquad  $1^{ - - }$ \qquad & \qquad  $3^{- - }$ \qquad & \qquad  $5^{- - }$ \qquad & \qquad  $7^{- - }$ \qquad \\
\hline \hline
 Relativistic many body \cite{LlanesEstrada:2005jf}             
&  3.95 &  4.15 &  5.05 &  5.90       \\ \hline
 Non-Relativistic constituent \cite{LlanesEstrada:2005jf}       
&  3.49 &  3.92 &  5.15 &  6.14       \\ \hline
 Wilson loop \cite{Kaidalov:1999yd}                             
&  3.49 &  4.03 &      &              \\ \hline
 Vacuum correlator \cite{Kaidalov:2005kz}                       
&  3.02 &  3.49 &  4.18 & 4.96          \\ \hline
 Vacuum correlator \cite{Kaidalov:2005kz}                       
&  3.32 &  3.83 &  4.59 & 5.25         \\ \hline
 Semi-relativistic potential  \cite{Mathieu:2008pb}               
& 3.99 & 4.16 & 5.26 &              \\ \hline
Anisotropic lattice \cite{Chen:2005mg}                         
& 3.83 & 4.20 &      &               \\ \hline
 Isotropic lattice \cite{Meyer:2004jc, Meyer:2004gx}            
&  3.24 & 4.33 &      &               \\ \hline
\end{tabular}
\caption{Glueball masses for $J^{PC}$ states expressed in GeV, with odd $J$, achieved with non-holographic models from the literature.}
%\end{ruledtabular}
\label{oddlit}
\end{table}
%%%%%%%%%%%%%%%%%%%%%%%%%%%%%%%%%%%%%%%
%%%%%%%%%%%%%%%%%%

\section{Hadronic Spectra for scalar mesons}\label{meson}

Mesons are bound states between a quark and an antiquark that can be represented by a spin singlet with total spin $S=0$ or a spin triplet with total spin $S=1$. The coupling between $S$ and the orbital angular momentum $L$ must be considered, producing a total angular momentum $J = L$ in the case of the singlet state, and $J = L - 1, L, L + 1$ in the case of the triplet state.

In mesonic spectroscopy \cite{Godfrey:1998pd}, mesons are characterized by $I^G(J^{PC})$, where $I$ is the isospin, $G$ is the $G$-parity defined $G=(-1)^I = \pm 1$, and $P$ is the $P$-parity defined for mesons as $P=(-1)^{L+1}$. Finally, $C$ is the $C$-parity defined as $C=(-1)^{L+S}$. In the boundary theory scalar mesons are represented by the operator:
\begin{equation}
{\cal O}_{SM}= \bar{q}\, D_{\lbrace J_1 \cdots} D_{J_m \rbrace}q \;\;\; {\rm with} \;\;\;\displaystyle\sum_{i=1} J_i = J ,
\end{equation}
\noindent where $J$ is the total angular momentum.

In this section we address light scalar mesons, i.e., $J=0$ and unflavored $(S = C = B = 0)$. 

Within the holographic approach, the description of the scalar glueball $(gg)$ and the scalar meson ($q \bar{q}$) is the same; however, the main difference is provided by the bulk mass, which defines the hadron identity. To study the scalar meson, we must to start from the action for a massive scalar field \eqref{esc_sw}, which will leads us to the ``Schr\"odinger-like'' equation \eqref{eq_4}.

\subsection{Results for scalar mesons spectra}\label{smres}

Employing the relationship $M^2_5 = (\Delta - p) (\Delta + p - 4)$, and identifying $M_5$ as the scalar meson bulk mass, the index of the $p-$form with the total angular momentum ($p=J=0$) for the scalar meson and $\Delta$ depicts the conformal dimension, which is $\Delta = 3$, as each quark contributes with  $3/2$. Finally, we rewrite Eq.\eqref{eq_4} with $M_5^2=-3$ as:
\begin{equation}\label{eq_sm}
- \psi''(z) + \left[ \frac{B'^2(z)}{4}  - \frac{B''(z)}{2} -3\; e^{\frac{-2 B(z)}{3}} \right] \psi(z) = - q^2 \psi(z), 
\end{equation}
\noindent where $B(z) = - 3 A(z)$. Solving \eqref{eq_sm} numerically with the warp factor constant $k$ identified as $k_{\rm sm}=-0.332^2$ GeV$^2$, we obtain the masses compatible with the family of the scalar meson $f_0$, with $I^GJ^{PC} = 0^+(0^{++})$, as indicated in table \ref{t3}. The error presented in last column of Table \ref{t3} ($\% M$) is the error defined by:  
\begin{equation}\label{M}
 \% M= \sqrt{ \left( \frac {\delta O_i}{O_i} \right)^2} \times 100,
\end{equation}
\noindent where $\delta O_i$ depicts the deviations between the data ($M_{\rm exp}$) and the model prediction ($M_{\rm th}$). Throughout the text, in the cases where the experimental is provided at intervals, as the $f_0(1370)$ state, we use the average value of the interval to evaluate the deviations. We moreover compute the total r.m.s error defined by:  
\begin{equation}\label{rms}
 \delta_{rms}= \sqrt{ \frac 1{N-N_p} \sum_{i=1}^N \left( \frac {\delta O_i}{O_i} \right)^2}  \times 100\,, 
\end{equation}
where $N$ and $N_p$ are the number of measurements and parameters, respectively. From Eq. \eqref{rms} we find that  $\delta_{rms} = 3.77 \%$ for table \ref{t3}.

%%%%%%%%%%%%%%%%%%%%%%%%%%
\begin{table}
%\begin{ruledtabular}
%\vspace{0.5 cm}
\centering
\begin{tabular}{|c|c|c|c|c|}
\hline
 &  \multicolumn{4}{c|}{Scalar meson $f_0$ ($0^+(0^{++})$)}  \\  
\cline{2-5}
 & $f_0$ meson & $M_{\rm exp}$ GeV \cite{Tanabashi:2018oca}& $M_{\rm th}$ GeV & $\% M$  \\
\hline \hline
\quad\, $n=1$ \quad \, & \, $f_0 (980)$\, &\, $0.990 \pm 0.02$   \,&\, 1.089 \,& \, 9.97 \, \\ \hline
$n=2$&\, $f_0 (1370)$\, &\, $1.2 \text{ to } 1.5$   \,&\,1.343 \,& \, 0.54 \, \\ \hline
$n=3$&\, $f_0 (1500)$\, &\, $1.504 \pm 0.006$ \,&\, 1.562 \,& \,3.87 \, \\ \hline
$n=4$&\, $f_0 (1710)$\, &\, $1.723^{+0.006}_{-0.005}$ \,&\, 1.757 \,& \, 1.96 \, \\ \hline
$n=5$&\, $f_0 (2020)$\, &\, $1.992 \pm 0.016$ \,&\, 1.933 \,& \, 2.96 \, \\ \hline
$n=6$&\, $f_0 (2100)$\, &\, $2.101 \pm 0.007$ \,&\, 2.095 \,& \, 0.27 \, \\ \hline
$n=7$&\, $f_0 (2200)$\, &\, $2.189 \pm 0.013$ \,&\, 2.246 \,& \,2.61 \, \\ \hline
$n=8$&\, $f_0 (2330)$\, &\, $2.337 \pm 0.014$ \,&\, 2.388 \,& \, 2.17 \, \\ \hline
\end{tabular}
\caption{Masses of light unflavored scalar meson $f_0\,(S = C = B = 0)$. Column $n=1,\, 2,\, 3, \cdots$ represents holographic radial excitation of scalar mesons. The ground state is represented by $n=1$. Column $M_{\rm exp}$ represents the experimental data from PDG \cite{Tanabashi:2018oca}. Column $M_{\rm th}$ represents masses obtained within the deformed $AdS_5$ space approach using Eq.\eqref{eq_sm} with $k_{\rm sm}= -0.332^2$ GeV$^2$. Column $\% M$ represents the error of $M_{\rm th}$ with respect to $M_{\rm exp}$, according to Eq. \eqref{M}.}
%\end{ruledtabular}
\label{t3}
\end{table}

\begin{figure}
	\centering
	\includegraphics[scale=0.75]{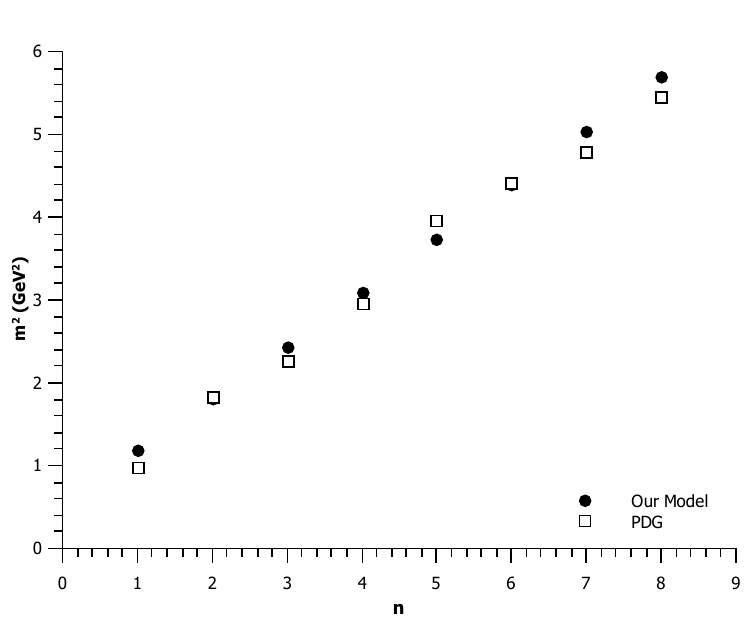}
	\caption{Scalar meson $f_0$ family squared masses as a function of their holographic radial excitation $n$ obtained within the deformed $AdS_5$ space approach (dots) and from PDG (squares), as presented in Table \ref{t3}.}
	\label{sm}
\end{figure}

Using the data from Table \ref{t3},  we plotted a Chew-Frautschi plane represented as $ n \times m^2$, where $n$ is the holographic radial excitation and, $m^2$ is the squared scalar meson mass represented by the dots (our model) or squares (PDG) in figure \ref{sm}. Using a standard linear regression method we obtain the experimental and theoretical Regge trajectories for the scalar meson $f_0$ family, such that:
\begin{equation}\label{RE_sm980_1}
m^2_{Exp} = (0.639 \pm 0.027) \; n + (0.458 \pm 0.135)\,, 
\end{equation}
\begin{equation}\label{RT_sm980_1}
m^2_{th}= (0.647 \pm 0.002) \; n + (0.513 \pm 0.011)\,. 
\end{equation}

The authors of Refs. \cite{Gherghetta:2009ac, Kelley:2011ds} within a holographic softwall model likewise computed the masses for the $f_0$ meson family and derived its Regge trajectory slightly differently from Eq. (\ref{RT_sm980_1}). This can be explained, as the data selection scenarios in these references are different from the current study. In these past studies, the scalar meson $f_0(500)$ was included, which might have caused the slight difference of the slope and the intercept compared to our study. 

To connect our results with the mesonic spectroscopy data 
\cite{Godfrey:1998pd, Anisovich:2000kxa, Ebert:2009ub, Chen:2018bbr} we split the isoscalar states $f_0$ into two sets. The first set, i.e., set 1, is related to the $ n \bar{n} = (u \bar{u} + d \bar{d})/\sqrt{2}$ states, which are represented by $f_0 (980)$, $f_0 (1500)$, $f_0 (2020)$ and $f_0 (2200)$. 
The second set, i.e., set 2, is related to $s \bar{s}$ states, also called $f'_0$, which is represented by $f_0(1370)$, $f_0(1710)$, $f_0(2100)$ and $f_0(2330)$.

\begin{figure}
	\centering
	\includegraphics[scale=0.75]{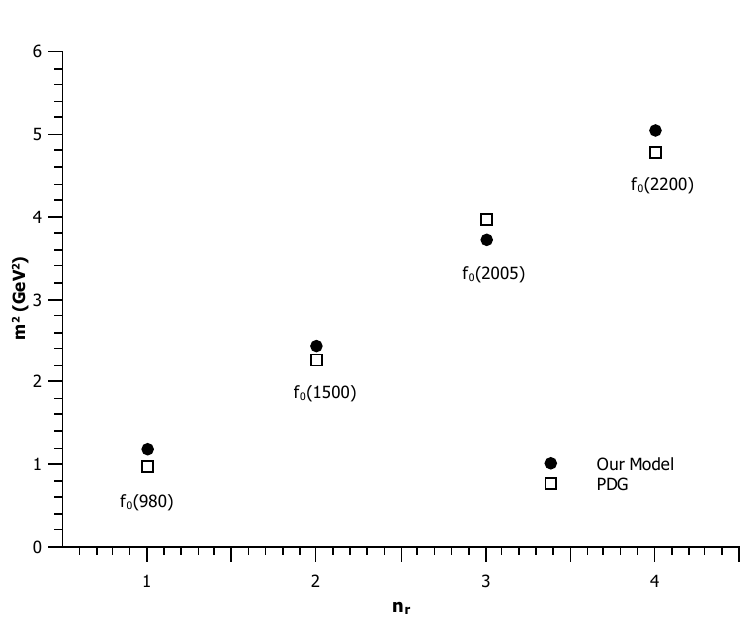}
	\caption{Scalar meson $f_0$ $ [n \bar{n} = (u \bar{u} + d \bar{d})/\sqrt{2}]$ states belonging to set 1 squared masses as a function of their spectroscopy radial excitation $n_r$ obtained within the deformed $AdS_5$ space approach (dots) and coming from PDG (squares).}
	\label{f01}
\end{figure}

Using the states that belong to set 1 we plot a Chew-Frautschi plane represented as $ n_r \times m^2$, where $n_r$ is the spectroscopy radial excitation and $m^2$ is the squared scalar meson mass represented by the dots (our model) or squares (PDG) in figure \ref{f01}. 
Using  a standard linear regression method we obtain the experimental and theoretical Regge trajectories for set 1, given by:
\begin{equation}\label{}
m^2_{Exp}= (1.314 \pm 0.017) \; n_r - (0.285 \pm 0.332)\,, 
\end{equation}
\begin{equation}\label{set1}
m^2_{th} = (1.288 \pm 0.009) \; n_r - (0.117 \pm 0.024)\,. 
\end{equation}

\begin{figure}
	\centering
	\includegraphics[scale=0.75]{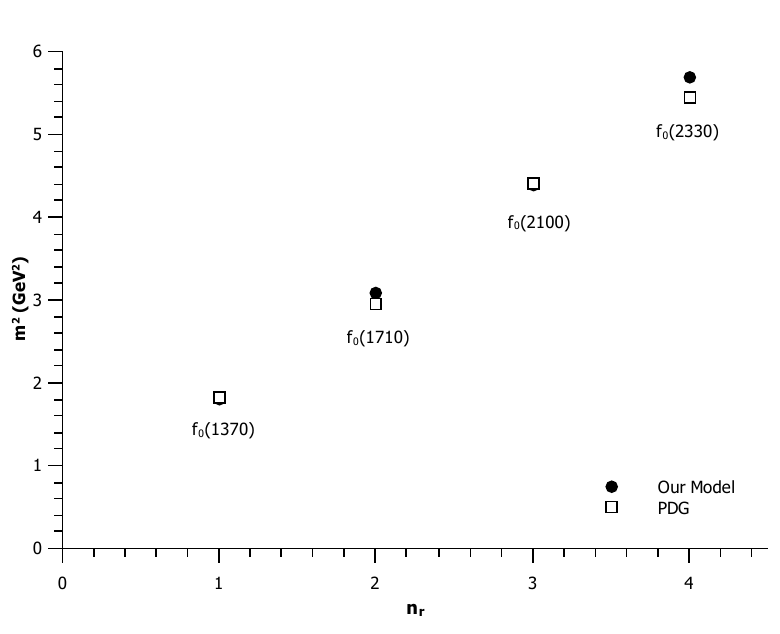}
	\caption{Scalar meson $f_0 [s \bar{s}]$ states belonging to set 2 squared masses as a function of their spectroscopy radial excitation $n_r$, obtained within the deformed $AdS_5$ space approach (dots) and from PDG (squares).}
	\label{f02}
\end{figure}

For the states  belonging to set 2, we plot Fig. \ref{f02} and obtain the experimental and theoretical Regge trajectories, given by:
\begin{equation}\label{}
m^2_{Exp}= (1.236 \pm 0.052) \; n_r - (0.576 \pm 0.142)\,, 
\end{equation}
\begin{equation}\label{set2} 
m^2_{th} = (1.300 \pm 0.005) \; n_r - (0.496 \pm 0.012)\,. 
\end{equation}

The Regge trajectories for scalar mesons belonging to the set 1 and 2 from our model, represented by Eqs. \eqref{set1} and \eqref{set2}, present Regge slopes ranged within the $1.25\pm 0.15$  GeV$^2$ which is close to the 
universal value $1.1$ GeV$^2$
 \cite{Anisovich:2000kxa, Iachello:1991re}.

%

%\clearpage

\section{Hadronic Spectra for vector mesons}\label{sec4}

Vector mesons have the same internal structure $(q \bar{q})$ as the scalar mesons, but with total angular momentum $J=1$. They are represented on the boundary theory by the operator:
\begin{equation}
{\cal O}_{VM}= \bar{q}\, \gamma^{\mu} D_{\lbrace J_1 \cdots} D_{J_m \rbrace}q \;\;\; {\rm with} \;\;\; \displaystyle\sum_{i=1} J_i = J\,. 
\end{equation}

In the holographic description, vector mesons are dual to the massive vector field in the $AdS_5$. Hence, the action for a massive vector field is needed, given by:
\begin{equation}\label{vec_sw}
S = -\frac{1}{2}\int d^5 x \sqrt{-g}\; [ \frac{1}{2} g^{pm} g^{qn} F_{mn} F_{pq} +M_5^2 g^{pm} A_p A_m],
\end{equation} 
\noindent where the vector field stress tensor is defined as $F_{mn} = \partial_m A_n - \partial_n A_m$.

The equations of motion are achieved by $\delta S / \delta A_n = 0$, so that:

\begin{equation}\label{eom_vec_4}
\partial_z[e^{-B(z)} F_{zn} \eta^{nq}] + \partial_{\mu}[e^{-B(z)} \eta^{m \mu} F_{m n} \eta^{nq}] - e^{-3B(z)} M_5^2 A_n \;\eta^{nq}= 0,
\end{equation}
\noindent where $B(z) = - A(z)$.

Considering a plane wave ansatz with the amplitude only depending on the $z$ coordinate and propagating in the transverse coordinates $x^{\mu}$ with momentum $q_{\mu}$, we obtain
\begin{equation}\label{anvec}
A_{\nu} (z, x^{\mu}) = v(z) e^{i q_{\mu} x^{\mu}} \epsilon_{\nu}, 
\end{equation}
\noindent assuming $A_z = 0$ and $\epsilon^{\nu} \epsilon_{\nu} = \eta^{\nu \lambda} \epsilon_{\nu} \epsilon_{\lambda}=1$ is the unitary $4-$vector defined in the transverse space to the $z$ coordinate, with components $\epsilon_{\nu} = 1/2(1,1,1,1)$.  We use the fact  $\partial_{\mu} A^{\mu} = 0$ which implies $q^{\mu} \epsilon_{\mu} = \eta^{\mu \lambda} q^{\mu} \epsilon_{\lambda} = q \cdot \epsilon = 0$ ensuring that the field can be written as a plane wave. Notably that $F_{zn} = \partial_z A_n$ and $\eta^{m \mu} \partial_{\mu}  F_{m n} = -q^2 A_n$. After some algebraic manipulation and defining $v(z) = \psi (z) e^{\frac{B(z)}{2}}$, we obtain the `Schr\"odinger-like'' equation, given by:
\begin{equation}\label{eom_vec_6}
- \psi''(z) + \left[ \frac{B'^2(z)}{4}  - \frac{B''(z)}{2} + e^{-2 B(z)} M^2_5 \right] \psi(z) = - q^2 \psi(z),
\end{equation}
\noindent where $E = -q^2$ are eigenenergies.

\subsection{Results for vector mesons spectra}\label{subvec}

We consider the case $J=1$. Then, recalling that $M^2_5 = (\Delta - p) (\Delta + p - 4)$, and identifying $M_5$ as the vector meson bulk mass, the index of $p-$form as total angular momentum ($p=J=1$) for the vector meson and $\Delta$ as the conformal dimension, which is $\Delta = 3$ as each quark contributes with  $3/2$. Finally, we rewrite Eq.\eqref{eom_vec_6} as: 
\begin{equation}\label{eq_vm}
- \psi''(z) + \left[ \frac{B'^2(z)}{4}  - \frac{B''(z)}{2} \right] \psi(z) = - q^2 \psi(z), 
\end{equation}
\noindent with $B(z) = - A(z)$ and $M^2_5=0$ for vector mesons.

Solving Eq. \eqref{eq_vm} numerically with the warp factor constant $k$ given by $k_{\rm vm}=-0.613^2$ GeV$^2$, we obtain the masses compatible with the family of vector meson $\rho$, with $I^GJ^{PC} = 1^+(1^{--})$, as indicated in Table \ref{t4}. The error presented in the last column of Table \ref{t4} ($\% M$) was definied in Eq.\eqref{M}. \noindent We also compute the total r.m.s error defined by Eq. \eqref{rms}. For table \ref{t4} we obtain $\delta_{rms} = 7.87 \%$.  

%%%%%%%%%%%%%%%%%%%%%
\begin{table}
%\begin{ruledtabular}
%\vspace{0.5 cm}
\centering
\begin{tabular}{|c|c|c|c|c|}
\hline
 &  \multicolumn{4}{c|}{Vector meson $\rho\;  (1^+(1^{--}$))}  \\  
\cline{2-5}
 & $\rho$ meson & $M_{\rm exp}$ GeV \cite{Tanabashi:2018oca}& $M_{\rm th}$ GeV & $\% M$  \\
\hline \hline
\qquad\, $n=1$ \qquad \, 
&\, $\rho (770)$\, &\, $0.77526 \pm 0.00025$ \,&\, 0.868327 \,& \, 12.0422 \, \\ \hline
$n=2$&\, $\rho (1450)$\, &\, $1.465 \pm 0.025$ \,&\, 1.228 \,& \,16.1775 \, \\ \hline
$n=3$&\, $\rho (1570)$\, &\, $1.570 \pm 0.070$ \,&\, 1.50399 \,& \, 4.20467 \, \\ \hline
$n=4$&\, $\rho (1700)$\, &\, $1.720 \pm 0.020$ \,&\, 1.73665 \,& \,0.968271 \, \\ \hline
$n=5$&\, $\rho (1900)$\, &\, $1.909 \pm 0.042$ \,&\, 1.94164 \,& \, 1.70972 \, \\ \hline
$n=6$&\, $\rho (2150)$\, &\, $2.155 \pm 0.021$ \,&\, 2.12696 \,& \, 1.30123 \, \\ \hline
\end{tabular}
\caption{Masses of light unflavored vector meson $\rho\,$ $(S = C = B = 0)$ . Column $n=1,\, 2,\, 3, \cdots$ represents holographic radial excitation of the vector mesons. The ground state is represented by $n=1$. Column $M_{\rm exp}$ represents experimental data from PDG \cite{Tanabashi:2018oca}. Column $M_{\rm th}$ represents masses obtained within the deformed $AdS_5$ space approach and using Eq.\eqref{eq_vm} with $k_{\rm vm}= -0.613^2$ GeV$^2$. Column $\% M$ represents the error of $M_{\rm th}$ with respect to $M_{\rm exp}$, according to Eq. \eqref{M}.}
%\end{ruledtabular}
\label{t4}
\end{table}
%%%%%%%%%%%%%%%%%%%%%%%%%%%%%%%%%%

\begin{figure}
	\centering
	\includegraphics[scale=0.75]{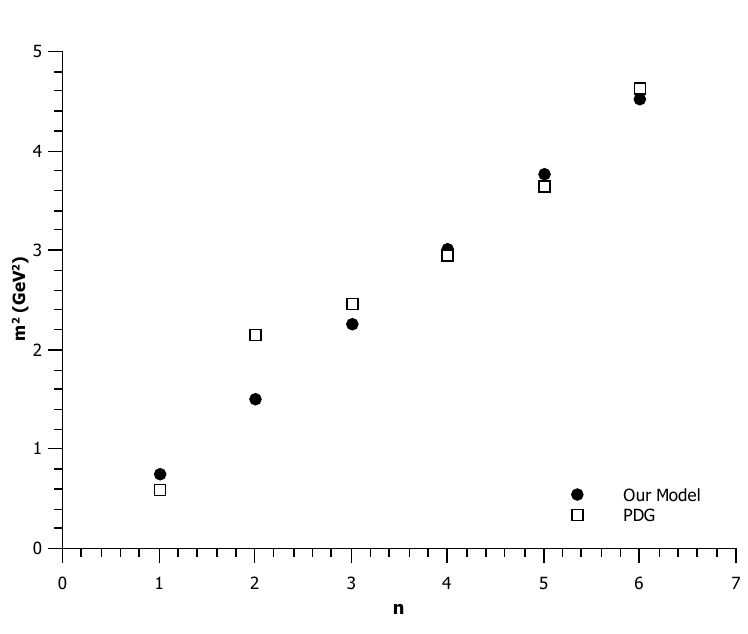}
	\caption{Vector meson $\rho$ family squared masses as a function of their holographic radial excitation obtained within the deformed $AdS_5$ space approach (dots) and from PDG (squares), as presented in Table \ref{t4}.}
	\label{vm}
\end{figure}

Using the data from Table \ref{t4} we plotted a Chew-Frautschi plane represented as $ n \times m^2$, where $n$ is the holographic radial excitation, and $m^2$ is the squared vector meson mass represented by the dots (our model) or squares (PDG) in Fig. \ref{vm}. Using a standard linear regression method we obtain the experimental and theoretical Regge trajectories for vector meson $\rho$, such that: 
\begin{equation}
m^2_{Exp} = (0.720 \pm 0.076) \;n  - (0.223 \pm 0.302)\,,
\end{equation}
\begin{equation}\label{RT_vm}
m^2_{th} = (0.754 \pm  8 \times 10^{-7}) \; n\,. 
\end{equation}

\noindent We did not include the intercept in Eq. \eqref{RT_vm}, because its value is very close to zero $(\approx 10^{-18})$. Moreover, in Eq. \eqref{RT_vm}, the uncertainty in the slope is very small indicating that this fit is practically a straight line.

The authors of Refs. \cite{Gherghetta:2009ac, Kelley:2011ds} also computed the masses for the $\rho$ meson family and derived their Regge trajectories within their holographic softwall model, obtaining approximately the same value for the slope and intercept (considering uncertainties) as the present study, Eq. (\ref{RT_vm}). The data select scenarios in those studies are different from the present study, as they included the vector meson $\rho(1282)$ as the first radial excited state and excluded the vector meson $\rho(1570)$, which the authors argue may be an OZI violating decay of the $\rho(1700)$. If we assume the existence of the $\rho(1282)$ as the first radial excitation $(n=2)$ of the $\rho$ meson family, then the corresponding percentage error $\%M$ in Table \ref{t4} would be smaller and so would the $\delta_{rms}$ error.

%\clearpage

\begin{figure}
	\centering
	\includegraphics[scale=0.75]{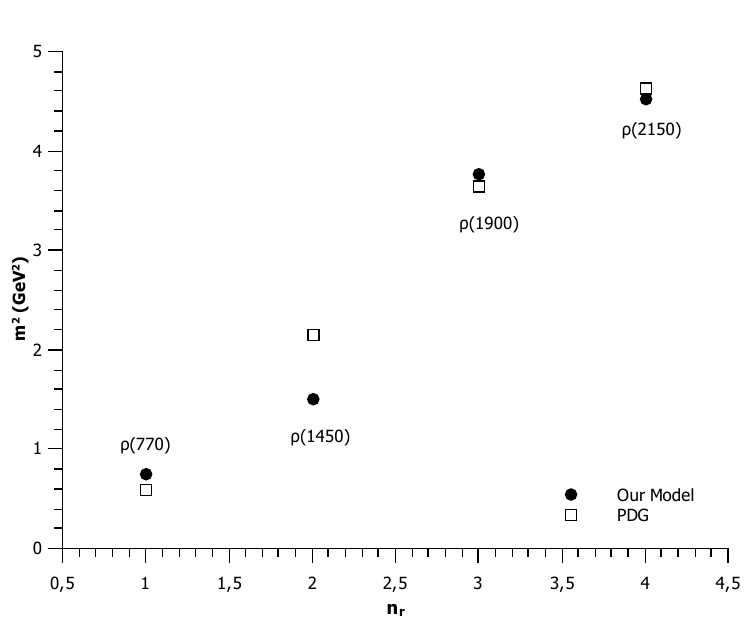}
	\caption{Vector mesons $\rho$ belonging to $S-$wave squared masses as a function of their spectroscopy radial excitation $n_r$ obtained within the deformed $AdS_5$ space approach (dots) and from PDG (squares).}
	\label{rho}
\end{figure}

As performed for the scalar mesons, we can resort to the mesonic spectroscopy data 
\cite{Godfrey:1998pd, Anisovich:2000kxa, Ebert:2009ub, Chen:2018bbr} and note that all vector mesons listed in Table \ref{t4} are not in the same spectroscopic state, meaning that only $\rho (770)$, $\rho (1450)$, $\rho (1900)$ and $\rho (2150)$ belong to the $S-$wave represented by $1^3S_1$, $2^3S_1$, $3^3S_1$ and $4^3S_1$, respectively. In this study, we used the spectroscopic notation, such as, $n_r^{2S+1}L_J$, where $n_r$ is the spectroscopy radial excitation. Using these states we plot in a Chew-Frautschi plane represented as $ n_r \times m^2$, where $n_r$ is the spectroscopy radial excitation and $m^2$ is the squared vector meson mass represented by the dots (our model) or squares (PDG) in figure \ref{rho}. 
Using a standard linear regression method, we obtain the experimental and theoretical Regge trajectories for vector meson $\rho$ belonging to the $S-$wave, so that:
\begin{equation}
m^2_{Exp} = (1.363 \pm 0.092) \;n_r  - (0.648 \pm 0.252)\,, 
\end{equation}
\begin{equation}\label{S-wave}
m^2_{th} = (1.357 \pm 0.213) \; n_r - (0.754 \pm 0.584)\,. 
\end{equation}

The Regge trajectory for vector mesons belonging to the $S-$wave from our model, represented by Eq. \eqref{S-wave}, yield a Regge slope in the range  $1.25\pm 0.15$  GeV$^2$ which is close to the universal value $1.1$ GeV$^2$ \cite{Anisovich:2000kxa, Iachello:1991re}. 

%\clearpage

Furthermore, if we follow the original motivation for the softwall model, it would be natural to suppose that $k_{sm}$ and $k_{vm}$ are related to the string tension for the flux tube that connects the two quarks inside the meson. This information is contained in the confining part of the $q\,\bar{q}$ potential, and it is  in principle a spin independent term. Therefore, in the AdS/QCD models with dilatons in the action, the slope parameter should be universal for scalar and vector mesons, as it happens in the conventional softwall model \cite{Karch:2006pv,Colangelo:2007pt}. 

Interestingly $k_{sm}$ and $k_{vm}$ are related, namely  $3k_{sm} \approx k_{vm}$.  This peculiarity could be attributed to the fact that in the EOM for scalar mesons, Eq.\eqref{eom_esc_2}, we performed the substitution $B(z) = -3 A(z)$. On the other hand, in the EOM for vector mesons, Eq.\eqref{eom_vec_4}, we used $B(z) = - A(z)$, leading to $k_{vm} \approx 3 k_{sm}$.

%\clearpage
\section{Hadronic spectra for baryons}\label{hf}

Within the quark model, constituent baryons are particles with a semi integer spin formed by a bound state of three valence quarks. In this study, we disregard states of baryons with higher complexity, composed of three quarks added to any number of quark and antiquark pairs, e.g., pentaquark states $(qqqq\bar{q})$. Hence, we use the following description for baryons, such that: 
\begin{equation}
|qqq \rangle_A = |{\rm color}\rangle_A \otimes |{\rm space; spin{\rm -}flavor}\rangle_S \,.
\end{equation}

The three colors are represented by an $SU(3)$ singlet, without dynamics and completely antisymmetric. The spatial wave function is related to $O(6)$, and the spin-flavor wave function is related to $SU(6)$. A review on baryon physics is provided in Refs. \cite{Klempt:2009pi, Klempt:2002cu}. In this study we are interested in light baryons composed of $u$ and $d$ quarks with a spin of 1/2 and with higher spins (3/2 and 5/2).

Within the holographic description, baryons are dual to the massive spinor fields in $AdS_5$. We start our discussion  from the free spinor field action without surface terms \cite{Henningson:1998cd, Mueck:1998iz, Abidin:2009hr, Gao:2010qk}:
\begin{equation}\label{acaofermions}
S =  \int_{AdS} d^{5} x\sqrt{g} \; \bar{\Psi}({\slashed D} - m_5 ) \Psi.
\end{equation}

We disregarded the hypersphere $S^5$, as for our purposes, the spinor field does not depend on these coordinates. Further, in the action \eqref{acaofermions}, $g$ is the determinant of the metric of the deformed $AdS_5$ space, given by Eq. \eqref{gs}.

As we deal with fermions in a curved space, we need to construct a local Lorentz frame or a vielbein. To simplify our notation, we will use $a, b, c$ to denote indexes in flat space, and $m, n, p, q$ to denote indexes in curved space (deformed $AdS_5$ space). The Greek indexes $\mu, \nu$ are defined in the Minkowski space. Thus, a useful choice is:
\begin{equation}\label{tetra}
e^{a}_m = e^{A(z)} \delta^{a}_m, \; \;\; e^{m}_a = e^{-A(z)} \delta^m_{a} e^{ma} = e^{-A(z)} \eta^{ma}, \; \; \; {\rm with} \; \;\;\;\;m = 0, 1, 2, 3, 5.
\end{equation}

The Levi-Civita connection is defined as:
\begin{equation}
\Gamma_{m n}^p = \frac{1}{2} g^{pq}(\partial_n g_{mq} + \partial_m g_{nq} - \partial_q g_{mn}), \;\;\; {\rm with} \;\;\;g_{mn} = e^{2 A(z)} \eta_{mn}.
\end{equation}

The corresponding spin connection $\omega^{\mu \nu}_{m}$, is given by:
\begin{equation}
\omega^{a b}_{m}  = e^a_n \partial_m e^{nb} + e^a_n e^{pb} \Gamma^n_{pm}.
\end{equation}
%¨
Because the only non-vanishing $\Gamma_{m n}^p $ are:
\begin{equation}\label{LV}
\Gamma_{\mu \nu}^5 = A'(z) \eta_{\mu \nu}, \; \; \Gamma_{5 5}^5 = -A'(z)  \; \; {\rm and} \; \; \Gamma_{\nu 5}^{\mu} = -A'(z) \delta^{\mu}_{\nu},
\end{equation}
\noindent we obtain:
\begin{equation}\label{spin}
\omega^{5 \nu}_{\mu} = - \omega^{\nu 5} _{\mu} = \partial_z A(z) \delta^{\nu}_{\mu},
\end{equation}
\noindent and all other components disappear.

The equations of motion are easily derived from Eq. \eqref{acaofermions}, so that:
\begin{equation}\label{fermioneom}
({\slashed D} - m_5 ) \Psi  = 0 \;\;\; {\rm and } \;\;\; \bar{\Psi} (-{\overleftarrow{\slashed D}} - m_5 )  = 0.
\end{equation}
Now using \eqref{gs}, \eqref{tetra} and \eqref{spin}, one can write the operator ${\slashed D}$ in \eqref{fermioneom}, so that:
\begin{equation}\label{slash}
{\slashed D} \equiv g^{mn} e^{a}_n \gamma_a \left( \partial_{m} + \frac{1}{2} \omega^{bc}_{m} \Sigma_{bc} \right) = e^{-A(z)} \gamma^5 \partial_5 + e^{-A(z)} \gamma^{\mu}  \partial_{\mu} + 2 A'(z)\gamma^5, 
\end{equation}
\noindent where we employed that $\gamma_a = (\gamma_{\mu}, \gamma_5)$, $\left\lbrace \gamma_a, \gamma_b \right\rbrace = 2 \eta_{ab} $, and $\Sigma_{\mu 5} = \frac{1}{4} \left[ \gamma_{\mu}, \gamma_5\right]$. Here, $\gamma_\mu$ are the usual Dirac's gamma matrices.

The first Dirac equation in Eq. \eqref{fermioneom} assumes the following form:
\begin{equation}\label{newdirac}
\left( e^{-A(z)} \gamma^5 \partial_5 + e^{-A(z)} \gamma^{\mu}  \partial_{\mu} + 2 A'(z)\gamma^5 - m_5\right) \Psi = 0, 
\end{equation}
\noindent where $\partial_5 \equiv \partial_z$,  $z$ is the holographic coordinate in the AdS space and $m_5$ is the fermion bulk mass. Considering a solution that can be decomposed into right- and left-handed chiral components, such as:
\begin{equation}\label{psi}
\Psi(x^{\mu}, z) = \left[ \frac{1 - \gamma^5}{2} f_L(z) + \frac{1 + \gamma^5}{2} f_R(z)\right] \Psi_{(4)}(x)\,, 
\end{equation}
\noindent with $\Psi_{(4)}(x)$ satisfying the Dirac equation $({\slashed D} - M)\Psi_{(4)}(x) = 0$ on the four-dimensional boundary space. The left and right modes also obey $\gamma^5 f_{L/R} = \mp f_{L/R}$ and $\gamma^{\mu}  \partial_{\mu} f_R = m f_L$ .

Because the Kaluza-Klein modes are dual to the chirality
spinors, we expand $\Psi_{L/R}$, so that:
\begin{equation}\label{kkmodes}
\Psi_{L/R} (x^{\mu}, z) = \sum_n f_{L/R}^n  (x^{\mu}) \phi_{L/R}^n (z).
\end{equation}

Using Eq. \eqref{kkmodes} with Eq. \eqref{psi} in Eq. \eqref{newdirac} we obtain a set with two coupled equations, such as:
\begin{equation}\label{mix1}
\left(\partial_z + 2 A'(z)\, e^{A(z)} + m_5\,e^{A(z)} \right) \phi_{L}^n (z) = + M_n \phi_{R}^n (z)
\end{equation}
\noindent and
\begin{equation}\label{mix2}
\left(\partial_z + 2 A'(z)\, e^{A(z)} - m_5\,e^{A(z)} \right) \phi_{R}^n (z) = - M_n \phi_{L}^n (z).
\end{equation}

Decoupling Eqs.\eqref{mix1} and \eqref{mix2}, and performing the following change of variables
\begin{equation}\label{cv}
\phi_{L/R}(z) = \psi(z) e^{-2 e^{A(z)}},
\end{equation}
\noindent we obtain a Schr\"odinger-like equation written for both right and left sectors, given by:
\begin{eqnarray}\label{scr}
-\psi_{R/L}''(z) + \left[ m_5^2 e^{2 A(z)} \pm m_5 e^{A(z)}A'(z) \right]\psi_{R/L}(z)  = M_n^2 \psi_{R/L}^n (z),
\end{eqnarray}
\noindent where $M_n$ in Eqs. \eqref{scr} depicts the four-dimensional fermion mass. 

\subsection{Results for spin 1/2 baryons spectra}

Here, we deal with light baryons with spin $S=1/2$ formed by $u$ and $d$ quarks. To this end, we consider the following operator on the boundary theory:
\begin{equation}\label{ob}
{\cal O}_{B}= q  D_{\lbrace \ell_1 \cdots} D_{\ell
_i }q D_{ \ell_{i+1} \cdots} D_{\ell_m \rbrace}q\;; \;\;\; {\rm with} \;\;\;\;\ \displaystyle\sum_{i=1} \ell_i = L,
\end{equation}
\noindent where $L$ is the orbital angular momentum. Here we consider only the case $L=0$.

From the AdS/CFT dictionary we find the following relationship for the fermion bulk mass $(m_5)$ and its conformal dimension ($\Delta$), so that:
\begin{equation}\label{massfermion}
|m_5|= \Delta - 2\,. 
\end{equation}

As each quark $u$ or $d$ contributes with $\Delta =3/2$, then the baryon formed by three quarks exhibits $\Delta = 9/2$ and consequently $m_5 = 5/2$.

Replacing $m_5 = 5/2$ in the Schr\"odinger-like equation \eqref{scr} and solving it numerically, with the warp factor constant $k$ identified as $k_{1/2}=0.205^2$ GeV$^2$, we obtain the masses compatible with the family of $N$ baryon, with $I(J^{P}) = 1/2(1/2^{+})$, as indicated in Table \ref{t5}. The error presented in last column of Table \ref{t5} ($\% M$) is defined in Eq.\eqref{M}. \noindent We also compute the total r.m.s error defined by Eq. \eqref{rms}. For Table \ref{t5} we obtain $\delta_{rms} = 4.09 \%$.

%%%%%%%%%%%%%%%%%%%%%
\begin{table}
%\begin{ruledtabular}
%\vspace{0.5 cm}
\centering
\begin{tabular}{|c|c|c|c|c|}
\hline
 &  \multicolumn{4}{c|}{Baryons $N (1/2^{+}$)}  \\  
\cline{2-5}
 & $N$ baryon & $M_{\rm exp}$ GeV \cite{Tanabashi:2018oca}& $M_{\rm th}$ GeV & $\% M$  \\
\hline \hline
\qquad \, $n=1$ \qquad \,                                    
&\, $N(939)$\, &\, $0.93949 \pm0.00005$ \,&\, 0.98683 \,& \, 5.04 \, \\ \hline
$n=2$&\, $N(1440)$\, &\, $1.360\text{ to }1.380\text{ }$ \,&\, 1.264 \,& \, 7.76 \, \\ \hline
$n=3$&\, $N(1710)$\, &\, $1.680\text{ to }1.720\text{ }$ \,&\, 1.531 \,& \, 9.94 \, \\ \hline
$n=4$&\, $N(1880)$\, &\, $1.820\text{ to }1.900\text{ }$ \,&\, 1.791 \,& \, 3.70 \, \\ \hline
$n=5$&\, $N(2100)$\, &\, $2.050\text{ to }2.150\text{ }$ \,&\, 2.046 \,& \, 2.58 \, \\ \hline
$n=6$&\, $N(2300)$\, &\, $2.300^{+0.006\;\;+0.1}_{-0.005\;\;-0}$ \,&\, 2.296 \,& \, 0.19 \, \\ \hline
\end{tabular}
\caption{Masses of $N(1/2^+)$ baryons. Column $n=1,\, 2,\, 3, \cdots$ represents holographic radial excitation. The ground state is represented by $n=1$. Column $M_{\rm exp}$ represents experimental data from PDG \cite{Tanabashi:2018oca}. Column $M_{\rm th}$ represents the masses of $N(1/2^+)$ baryons with $k_{1/2}= 0.205^2$ {\rm GeV}$^2$, obtained within the deformed $AdS_5$ space approach and using Eq.\eqref{scr}. Column $\% M$ represents the error of $M_{\rm th}$ with respect to $M_{\rm exp}$, according to Eq. \eqref{M}.}
%\end{ruledtabular}
\label{t5}
\end{table}
%%%%%%%%%%%%%%%%%%%%%%%%%%%%%%%%%%

\begin{figure}
	\centering
	\includegraphics[scale=0.75]{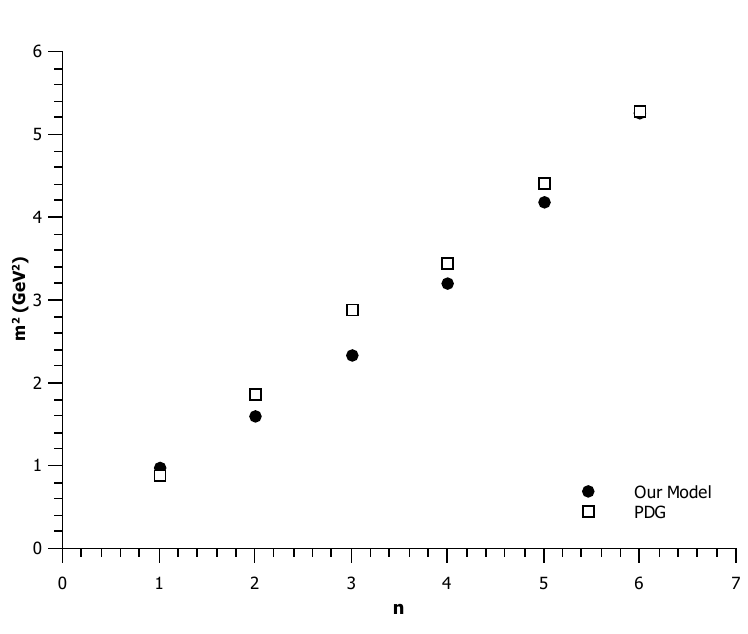}
	\caption{$N(1/2^+)$ baryon family squared masses as a function of their holographic radial excitation obtained within the deformed $AdS_5$ space approach (dots) and from PDG (squares), as presented in Table \ref{t5}.}
	\label{bar_1_2}
\end{figure}

Using the data from Table \ref{t5}, we plotted a Chew-Frautschi plane represented as $ n \times m^2 $, where $n$ is the holographic radial excitation and $m^2$ is the squared $N(1/2^+)$ baryon mass represented by the dots (our model) or squares (PDG) in Fig. \ref{bar_1_2}. Usig a standard linear regression method, we obtain the experimental and theoretical Regge trajectories for the $N(1/2^+)$ baryon, such that: 
\begin{equation}
m^2_{Exp} = (0.863 \pm 0.029) \;n  + (0.114 \pm 0.111)\,, 
\end{equation} 
\begin{equation}
m^2_{th} = (0.860 \pm  0.042) \; n - (0.081 \pm 0.164)\,. 
\end{equation}
%%%%%%%%%%
%\clearpage

\begin{figure}
	\centering
	\includegraphics[scale=0.75]{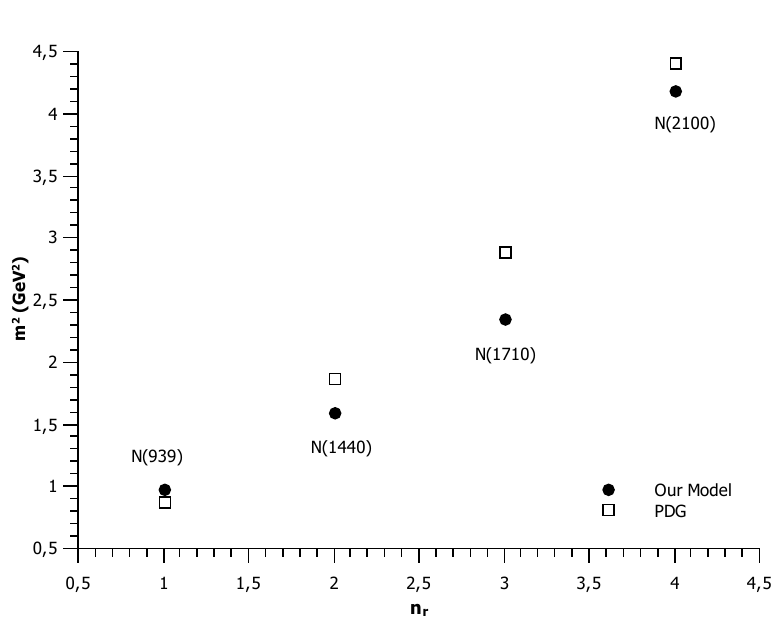}
	\caption{$N(1/2^+)$ baryons belonging to the $(56,^28)_0$ state squared masses as a function of their spectroscopy radial excitation $n_r$ obtained within the deformed $AdS_5$ space approach (dots) and from PDG (squares).}
	\label{bar_N}
\end{figure}

As performed for the scalar and vector mesons we resort to baryonic spectroscopy and attempt to recognize which baryons among those listed in Table \ref{t5} belong to the same spectroscopy state. According to Refs. \cite{Klempt:2009pi, Klempt:2002cu}, we see that the states $N(939)$, $N(1440)$, $N(1710)$ and $N(2100)$ belong to the state $D_L \equiv(56,^28)_0$ with spectroscopy radial excitation $n_r$, corresponding to $n_r = 1, 2, 3, 4$, respectively, with orbital angular momentum $L=0$. In this notation, $D$ represents the 56-plet, which can be broken into an octet with spin 1/2 ($^2 8$) and a decuplet with spin 3/2 ($^4{10}$). For these mentioned states, we plot a Chew-Frautschi plane represented as $ n_r \times m^2$, where $n_r$ is the spectroscopy radial excitation and $m^2$ is the squared $N(1/2^+)$ baryon mass belonging to the $(56,^28)_0$ state  represented by the dots (our model) or squares (PDG) in Fig. \ref{bar_N}. 
Using a standard linear regression method, we obtain the experimental and theoretical Regge trajectories for $N(1/2^+)$ baryon in the $(56,^28)_0$ state, so that:
\begin{equation}
m^2_{Exp} = (1.160 \pm 0.090) \;n_r  - (0.384 \pm 0.246)\,,
\end{equation}
\begin{equation}\label{1/2}
m^2_{th}= (1.038 \pm  0.204) \; n_r - (0.320 \pm 0.560)\,. 
\end{equation}

The Regge trajectory for the $N(1/2^+)$ baryon belonging to the same multiplet comes from our model, represented by Eq. \eqref{1/2}, and presents a Regge slope in the range $1.081 \pm 0.036$  GeV$^2$ which is close to the universal value $1.1$ GeV$^2$
 \cite{Klempt:2002vp}.

%

%\clearpage

\subsection{Results for higher spin baryons spectra}\label{hfhs}

Here,  we deal with light baryons, according to the same structure as in the previous section and a higher spin, meaning, e.g., $S=3/2$ or $S=5/2$. To this end, we we employ the same approach for the higher spin glueball as in Subsection \ref{sub}. To obtain the spectrum for spin $3/2$ baryons we insert symmetrized covariant derivatives in the operator ${\cal O}_{B}$, given by Eq. \eqref{ob}. Then, the conformal dimensions related to the spin $3/2$ baryons is now $\Delta_{3/2} = 11/2$, with $m_5 = 7/2$. Solving Eq. \eqref{scr}  with the warp factor constant $k$ given  by $k_{3/2}=0.205^2$ GeV$^2$, we obtain the masses compatible with the family of $N$ baryon, with $I(J^{P}) = 1/2(3/2^{+})$, as indicated in Table \ref{t6a}. The error presented in last column of table \ref{t6a} ($\% M$) is defined in Eq.\eqref{M}. \noindent We also compute the total r.m.s error defined by Eq. \eqref{rms}. For Table \ref{t6} one finds $\delta_{rms} = 9.00\%$.

%%%%%%%%%%%%%%%%%%%%%
\begin{table}
%\begin{ruledtabular}
%\vspace{0.5 cm}
\centering
\begin{tabular}{|c|c|c|c|c|}
\hline
 &  \multicolumn{4}{c|}{Baryons $N (3/2^{+}$)}  \\  
\cline{2-5}
 & $N$ baryon & $M_{\rm exp}$ GeV \cite{Tanabashi:2018oca}& $M_{\rm th}$ GeV & $\% M$  \\
\hline \hline
\qquad \, $n=1$ \qquad \,                                    
&\, $N(1720)$\, &\, $1.660\text{ to }1.690\text{ }$ \,&\,  1.326\,& \, 23.05\%  \, \\ \hline
$n=2$&\, $N(1900)$\, &\, $1.900\text{ to }1.940\text{ }$ \,&\, 1.606  \,& \,12.27\%  \, \\ \hline
$n=3$&\, $N(2040)$\, &\, $2.040^{+0.003}_{-0.004} \pm 0.025$ \,&\, 1.878 \,& \, 8.72\%  \, \\ \hline
\end{tabular}
\caption{Masses of $N(3/2^+)$ baryons. Column $n=1,\, 2,\, 3, \cdots$ represents holographic radial excitation. The ground state is represented by $n=1$. Column $M_{\rm exp}$ represents experimental data from PDG \cite{Tanabashi:2018oca}. Column $M_{\rm th}$ represents the masses of $N(3/2^+)$ baryons with $k_{3/2}= 0.205^2$ {\rm GeV}$^2$, obtained within the deformed $AdS_5$ space approach and using Eq.\eqref{scr}. Column $\% M$ represents the error of $M_{\rm th}$, according to Eq. \eqref{M}.}
%\end{ruledtabular}
\label{t6a}
\end{table}
%%%%%%%%%%%%%%%%%%%%%%%%%%%%%%%%%%

Observing the column $(\%M)$ in Table \ref{t6a}, The errors between $M_{\rm exp}$ and $M_{\rm th}$ are excessively high, especially for $n=1$ and $n=2$ states. 
A possible reinterpretation would be a missing state, which represents the ground state for the $N(3/2^+)$ baryons family. 
Taking into account this assumption, regarding a possible missing state, we can reinterpret Table \ref{t6a} as in Table \ref{t6}, where in the first line we present a possible baryon prediction obtained within the deformed AdS model. The error presented in last column of table \ref{t6} ($\% M$) is defined in Eq.\eqref{M}, We also compute the total r.m.s error defined by Eq. \eqref{rms}. For table \ref{t6} we find $\delta_{rms} = 2.13 \%$. We excluded our prediction of the errors calculation. The error in Table \ref{t6a} are greater than in Table \ref{t6}. 
However,this possible ground state $N(1330)$ that we are reinterpreting is not found in PDG. In PDG the $\Delta^ +$ states with $J^P=3/2^+$ and mass around 1320 MeV, is found; hence, our model is possibly not capable to distinguishing these two states. This first state could be $\Delta (1232)$ as both trajectories, $\Delta$ and $N(3/2^+)$, are supposed to be degenerate in the chiral limit, as it happens with mesons $\rho$
and $\omega$. 

%%%%%%%%%%%%%%%%%%%%%
\begin{table}
%\begin{ruledtabular}
%\vspace{0.5 cm}
\centering
\begin{tabular}{|c|c|c|c|c|}
\hline
 &  \multicolumn{4}{c|}{Baryons $N (3/2^{+}$)}  \\  
\cline{2-5}
 & $N$ baryon & $M_{\rm exp}$ GeV \cite{Tanabashi:2018oca}& $M_{\rm th}$ GeV & $\% M$  \\
\hline \hline
\qquad \, $n=1$ \qquad \,                                   
&\, \, &\,  \,&\, 1.326 \,& \,  \, \\ \hline
$n=2$                                   
&\, $N(1720)$\, &\, $1.660\text{ to }1.690\text{ }$ \,&\, 1.606 \,& \, 4.14 \, \\ \hline
$n=3$&\, $N(1900)$\, &\, $1.900\text{ to }1.940\text{ }$ \,&\, 1.878 \,& \, 2.19 \, \\ \hline
$n=4$&\, $N(2040)$\, &\, $2.040^{+0.003}_{-0.004} \pm 0.025$ \,&\, 2.144 \,& \, 5.09 \, \\ \hline
\end{tabular}
\caption{Masses of $N(3/2^+)$ baryons. Column $n=1,\, 2,\, 3, \cdots$ represents holographic radial excitation. The ground state is represented by $n=1$. Column $M_{\rm exp}$ represents experimental data from PDG \cite{Tanabashi:2018oca}. Column $M_{\rm th}$ represents the masses of $N(3/2^+)$ baryons with $k_{3/2}= 0.205^2$ {\rm GeV}$^2$, obtained within the deformed $AdS_5$ space approach and using Eq.\eqref{scr}. Column $\% M$ represents the error of $M_{\rm th}$ with respect to $M_{\rm exp}$, according to Eq. \eqref{M}. In the first line, we present a possible baryon prediction  within our model.}
%\end{ruledtabular}
\label{t6}
\end{table}
%%%%%%%%%%%%%%%%%%%%%%%%%%%%%%%%%%

\begin{figure}
	\centering
	\includegraphics[scale=0.75]{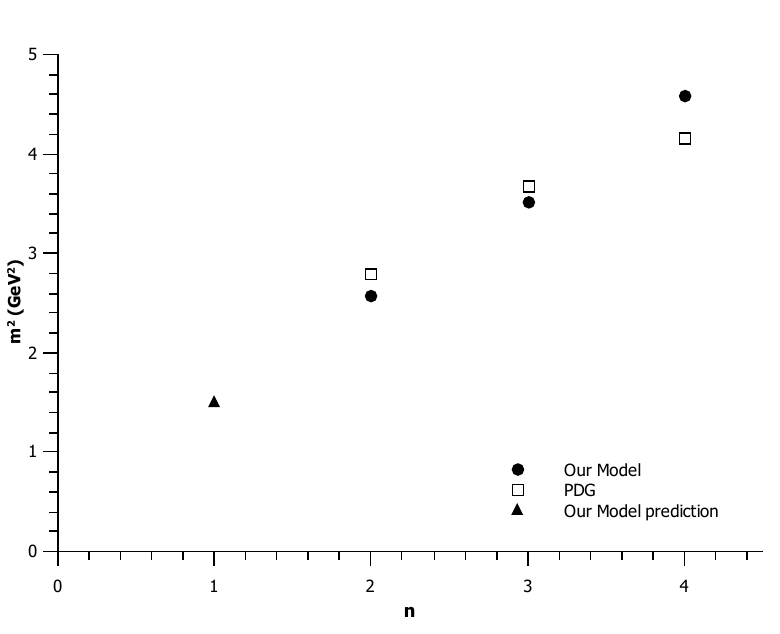}
	\caption{$N(3/2^+)$ baryon family squared masses as a function of their holographic radial excitation obtained within the deformed $AdS_5$ space approach (dots), our model prediction (triangle) and from PDG (squares), as presented in Table \ref{t6}.}
	\label{bar_3_2}
\end{figure}

Using the data from Table \ref{t6} we plotted a Chew-Frautschi plane represented as $ n \times m^2 $, where $n$ is the holographic radial excitation and $m^2$ is the squared $N(3/2^+)$ baryon mass represented by the dots (our model), by the triangle (our model prediction) or squares (PDG) in Fig. \ref{bar_3_2}. Using a standard linear regression method we obtain the experimental and theoretical Regge trajectories for $N(3/2^+)$ baryons, so that: 
\begin{equation}\label{RE_NB32}
m^2_{Exp} = (0.678 \pm 0.117) \;n  + (1.517 \pm 0.364)\,,
\end{equation}
\begin{equation}\label{RT_NB32}
 m^2_{th} = (1.021  \pm  0.017) \;n + (0.501 \pm 0.047)\,.
\end{equation}
For the linear fit in Eq.\eqref{RT_NB32} we took into account our predicted state.

The Regge trajectory for the $N(3/2^+)$ baryon family from our model, represented by Eq. \eqref{RT_NB32}, present a Regge slope in the range $1.081 \pm 0.036$  GeV$^2$ which is close to the universal value $1.1$ GeV$^2$
 \cite{Klempt:2002vp}. 

%%%%

At this point, we deal with baryons of spin 5/2. To this end, once again, insert one more symmetrized covariant derivative in the operator ${\cal O}_{B}$ given by Eq. \eqref{ob}. Then, we obtain the conformal dimension, given by $\Delta_{5/2} = 13/2$,  which provides $m_5 = 9/2$. Solving Eq. \eqref{scr} with the warp factor constant $k$ given by $k_{5/2}=0.190^2$ GeV$^2$, we obtain the masses compatible with the family of $N$ baryons, with $I(J^{P}) = 1/2(5/2^{+})$, as indicated in Table \ref{t7}. The error presented in the last column of table \ref{t7} ($\% M$) is defined in Eq. \eqref{M}. \noindent We compute the total r.m.s error defined by Eq. \eqref{rms}. For Table \ref{t7} one finds $\delta_{rms} = 2.76 \%$.

%%%%%%%%%%%%%%%%%%%%%
\begin{table}
%\begin{ruledtabular}
%\vspace{0.5 cm}
\centering
\begin{tabular}{|c|c|c|c|c|}
\hline
 &  \multicolumn{4}{c|}{Baryons $N (5/2^{+}$)}  \\  
\cline{2-5}
 & $N$ baryon & $M_{\rm exp}$ GeV \cite{Tanabashi:2018oca}& $M_{\rm th}$ GeV & $\% M$  \\
\hline \hline
\qquad \, $n=1$ \qquad \,                                    
&\, $N(1680)$\, &\, $1.665\text{ to }1.680\text{ }$ \,&\, 1.542 \,& \, 7.78 \, \\ \hline
$n=2$&\, $N(1860)$\, &\, $1.830^{+120}_{-60}$ \,&\, 1.804 \,& \,1.44 \, \\ \hline
$n=3$&\, $N(2000)$\, &\, $2.090 \pm 120$ \,&\, 2.059 \,& \,1.49 \, \\ \hline
\end{tabular}
\caption{Masses of $N(5/2^+)$ baryons. Column $n=1,\, 2,\, 3, \cdots$ represents holographic radial excitation. The ground state is represented by $n=1$. Column $M_{\rm exp}$ represents experimental data from PDG \cite{Tanabashi:2018oca}. Column $M_{\rm th}$ represents the masses of $N(5/2^+)$ baryons with $k_{5/2}= 0.190^2$ {\rm GeV}$^2$, obtained within the deformed $AdS_5$ space approach and using Eq.\eqref{scr}. Column $\% M$ represents the error of $M_{\rm th}$ with respect to $M_{\rm exp}$, according to Eq. \eqref{M}.}
%\end{ruledtabular}
\label{t7}
\end{table}
%%%%%%%%%%%%%%%%%%%%%%%%%%%%%%%%%%

\begin{figure}
	\centering
	\includegraphics[scale=0.75]{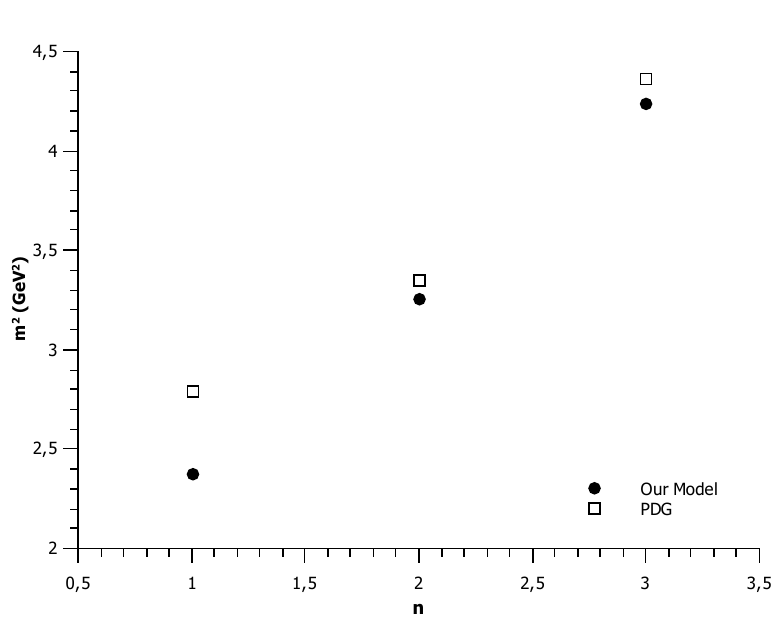}
	\caption{$N(5/2^+)$ baryon family squared masses as a function of their holographic radial excitation obtained within the deformed $AdS_5$ space approach (dots) and from PDG (squares), as presented in Table \ref{t7}.}
	\label{bar_5_2}
\end{figure}

From Table \ref{t7}, we plotted a Chew-Frautschi plane  as $ n \times m^2 $, where $n$ is the holographic radial excitation and $m^2$ is the squared $N(5/2^+)$ baryon mass represented by the dots (our model) or squares (PDG) in figure \ref{bar_5_2}. Using a standard linear regression method we obtain the experimental and theoretical Regge trajectories for $N(5/2^+)$ baryons, so that:
\begin{equation}\label{RE_NB52}
m^2_{Exp} = (0.785 \pm 0.135) \; n + (1.934 \pm 0.291)\,,
\end{equation}
\begin{equation}\label{RT_NB52}
m^2_{th} = (0.931 \pm  0.031) \;n  + (1.429 \pm 0.068)\,.
\end{equation}

The Regge trajectory for the $N(5/2^+)$ baryon family from our model, represented by Eq. \eqref{RT_NB52}, present a Regge slope near the range $1.081 \pm 0.036$  GeV$^2$ which is close to the universal value $1.1$ GeV$^2$
 \cite{Klempt:2002vp}.

%%%

Notably, the numeric values of the warp factor constant $k$ for the baryons in this study are approximately independent of their spin, meaning that $k_{1/2} = k_{3/2} \approx k_{5/2}$.

%\clearpage

%%%%%%%%%%%%%%%%%%%%
%%%%%%%%%%%%%%%%%%%%
\section{Summary and conclusions}\label{sec6}
%\section{Some title}
%Please always give a title also for appendices.

We studied the hadronic spectra based on the  holographic model within deformed $AdS_5$ space metrics, establishing that the warp factor is $A(z) = -\log(z) + {kz^2 }/{2}$ instead of $A(z) = -\log(z)$ of the pure AdS space. This deformation implies that there is no dilaton field in the action as in the original softwall model. In our model, different values are needed for the parameter $k$ for each particle sector.  A possible interpretation for this behavior is the following: if one assumes that the QCD vacuum is defined by the metric, our result of multiple values of $k$ indicates that the QCD vacuum should be non-trivial and possibly  composed of various non-equivalent vacua states.

The main achievement of this study is to provide an approach that can adequately accommodate the spectra for even and odd glueballs, scalar ($0^+(0^{++})$) and vector mesons $(1^+(1^{--}))$, as well as $N$ baryons with spin 1/2, 3/2 and 5/2 using the same holographic approach. This implies that the masses of these mentioned particles, computed using our model, and the derived Regge trajectories are in agreement with the literature. 

For the even and odd glueball cases, our model provides appropriate masses, as indicated in Tables \ref{t1} and \ref{t2}, when compared with other approaches (a summary of even and odd spin glueball masses obtained from lattice and other models, is given in Tables \ref{pomlit} and \ref{oddlit}). The computed masses for higher even and odd spin glueballs were placed in a Chew-Frautschi plane $ m^2 \times J $. We derived the Regge trajectories related to the pomeron and the odderon which are likewise in agreement with the literature.  

Our model performs well for scalar mesons, providing appropriate masses for the $f_0$ ($0^+(0^{++})$), as indicated in Table \ref{t3}, as compared with the data from PDG  \cite{Tanabashi:2018oca}. The obtained Regge trajectory from $m^2 \times n$ is compatible with the one of the holographic softwall model \cite{Gherghetta:2009ac, Kelley:2011ds}. Using spectroscopy data for the scalar mesons, we split them into two sets. The first one contains only  $n\bar n= 1/\sqrt2(u\bar u + d\bar d)$, while the second contains only $s\bar s$. For these sets, we derived Regge trajectories in $m^2 \times n_r$ and found that they are compatible with the literature   \cite{Anisovich:2000kxa, Iachello:1991re}. 

For the vector meson $\rho (1^+(1^{--}))$ our model provided appropriate masses as well, as shown in Table \ref{t4} compared with PDG. 
The obtained Regge trajectory from $m^2 \times n$ is compatible with the one from the holographic softwall model \cite{Gherghetta:2009ac, Kelley:2011ds}. 
Using the spectroscopy data for the vector mesons, we selected the $S-$wave states and derived their Regge trajectory in $m^2 \times n_r$ finding agreement with the literature  \cite{Anisovich:2000kxa, Iachello:1991re}. 
      
Our model also provides appropriate masses for the $N(1/2^+)$ baryon, as shown in Table \ref{t5}, compared with PDG. In this case, we likewise used the baryonic spectroscopic data to select states in the same multiplet, only varying their radial excitation. From these states we derived the Regge trajectory, which was compatible with the literature  \cite{Klempt:2002vp}.  

For the $N(3/2^+)$ baryon, we obtained unsatisfactory results for the masses as shown in Table \ref{t6a}. These results can be improved by introducing a hypothetical baryonic state to occupy the ground state (Table \ref{t6}). Using this assumption, the errors decrease and the derived Regge trajectory is compatible with the literature  \cite{Klempt:2002vp}. 

Finally, for the $N(5/2^+)$ baryon our model provides appropriate masses, as shonw in Table \ref{t7}, in comparison with PDG, and the Regge trajectory is in a reasonable agreement with the literature  \cite{Klempt:2002vp}.

It is important to note that in our model, the form of the warp factor is the same for all studied particles, whereas the parameter $k$ is adjusted for each case. In ref.  \cite{Forkel:2007cm} the authors employ different  warp factors for each kind of particle that is dependent on the angular momentum. In our case, for even and odd glueballs, the value of $k$ is the same, $k_{gbe}=k_{gbo}=0.31^2$ GeV$^2$. For scalar and vector mesons, we found that $k_{vm}\approx 3 k_{sm}$, as discussed at the end of Subsection \ref{subvec}. For the baryonic case, we found $k_{1/2}=k_{3/2}\approx k_{5/2}$. 

The Regge trajectories presented in this study related to hadronic spectroscopy for scalar mesons \eqref{set1}, \eqref{set2}, vector mesons \eqref{S-wave}, and baryons  \eqref{1/2}, \eqref{RT_NB32},  \eqref{RT_NB52} point towards a universal Regge slope around $1.1$ GeV$^2$ in accordance with the literature  \cite{Anisovich:2000kxa, Iachello:1991re, Klempt:2002vp, Bugg:2004xu}.

Our model finds different signs in the exponential of the warp factor, depending on each hadronic sector. Hence the question regarding the sign of the dilaton in the original softwall model persists. Despite the different signs for different sectors, we have no massless fields in our model. This a consequence of the deformed geometry instead of the introduction of a dilaton field in the action as in the original softwall model. 
   
\begin{acknowledgments} 
 The authors would like to thank Carlos Alfonso Ballon Bayona for useful discussions, and Oleg Andreev and Song He for useful correspondence. H.B-F. would like to thank partial financial support from Conselho Nacional de Desenvolvimento Cientifíco e Tecnológico (CNPq) and Coordenação de Aperfeiçoamento de Pessoal de Nível Superior (Capes) (Brazilian Agencies). A. V. and  M. A. M. C.  would like to thank the financial support given by FONDECYT (Chile) under Grants No. 1180753  and No. 3180592 respectively. D.L. is supported by the National Natural Science Foundation of China (11805084), the PhD Start-up Fund of Natural Science
Foundation of Guangdong Province (2018030310457) and Guangdong Pearl River Talents Plan (2017GC010480).
\end{acknowledgments}

\end{document}